\documentclass[prb,superscriptaddress,twocolumn, longbibliography]{revtex4-2}

\usepackage{amsmath,amssymb}
\usepackage{physics}
\usepackage{graphicx}
\usepackage{bm}
\usepackage{bbm}
\usepackage[colorlinks,bookmarks=false,citecolor=blue,linkcolor=black,urlcolor=red]{hyperref}
\usepackage{color} 
\usepackage{newtxtext,newtxmath}

\allowdisplaybreaks

\begin{document}

\title{Discrete time crystals detected by time-translation twist}

\date{\today}

\author{Ryota Nakai}
\affiliation{RIKEN Center for Quantum Computing (RQC), Wako, Saitama, 351-0198, Japan}
\affiliation{Department of Physics, Kyushu University, Fukuoka 819-0395, Japan}

\author{Taozhi Guo}
\affiliation{Department of Physics, Princeton University, Princeton, New Jersey 08544, USA}

\author{Shinsei Ryu}
\affiliation{Department of Physics, Princeton University, Princeton, New Jersey 08544, USA}

\begin{abstract}
We introduce a boundary condition twisted by time translation
as a novel probe to characterize dynamical phases in periodically driven (Floquet) quantum systems. 
Inspired by twisted boundary conditions in equilibrium systems, this approach modifies the temporal evolution of the system upon completing a spatial loop, enabling the identification of distinct Floquet phases, including discrete time crystals (DTCs).
By studying the spectral form factor (SFF) and its response to the
twist, we uncover signatures of time-crystalline order, which exhibits periodic dependence on the twist parameter analogous to the Little-Parks effect in superconductors. We apply this framework to the kicked Ising model, demonstrating that our
twist can distinguish time-crystalline phases. 
\end{abstract}

\maketitle

\section{Introduction}

Periodically driven quantum systems extend the concept of topological and spontaneously symmetry broken phases into dynamical ones that do not exist in static systems \cite{khemani2019briefhistorytimecrystals,annurev:/content/journals/10.1146/annurev-conmatphys-031218-013423,annurev:/content/journals/10.1146/annurev-conmatphys-031218-013721,annurev:/content/journals/10.1146/annurev-conmatphys-031119-050658, RevModPhys.95.031001}.
The examples include time crystals 
\cite{PhysRevLett.109.160401,
PhysRevA.91.033617,PhysRevLett.116.250401,PhysRevLett.117.090402, PhysRevLett.118.030401,*PhysRevLett.118.269901,Sacha_2018}
and Floquet topological phases, such as anomalous Floquet topological Anderson insulators
\cite{PhysRevB.79.081406,
  PhysRevB.82.235114,
  Lindner2011,  
  PhysRevLett.106.220402,
    annurev:/content/journals/10.1146/annurev-conmatphys-031218-013423,
    Rudner2020,
  annurev:/content/journals/10.1146/annurev-conmatphys-031218-013721}.
In recent years, Floquet engineering has been widely applied to a diverse range of condensed matter systems,  including graphene, topological insulators, monolayer transition-metal dichalcogenides
and others
\cite{McIver_2019,
Wang_2013, Mahmood_2016, Ito_2023,
Sie_2014, Kobayashi2023,
PhysRevLett.131.116401,
Reutzel2020,Bukov_2015},
as well as various synthetic platforms 
\cite{Peng2016,PhysRevA.89.063628,PhysRevLett.122.173901,
Wintersperger_2020},
to realize novel nonequilibrium phases of matter that do not exist in equilibrium. 
Topological Floquet systems have been experimentally realized in synthetic quantum systems
\cite{Peng2016,PhysRevA.89.063628,PhysRevLett.122.173901, 
Wintersperger_2020,
Braun_2024}.
Furthermore,
Floquet discrete time crystals (DTCs) in the many-body localized phase have been observed, e.g., in trapped ions \cite{Zhang2017}, spin systems \cite{Choi2017,doi:10.1126/science.abk0603}, and on quantum computers \cite{doi:10.1126/sciadv.abm7652}.
These advances have significantly broadened our understanding of dynamical phases, paving the way for new theoretical and experimental discoveries.

These new dynamical phases compel us to explore ways to characterize them. In this paper, we adopt the concept of twisting boundary conditions, a key tool in characterizing static phases of matter. Here, we extend this idea to the realm of non-equilibrium phases of matter.

Our extension is twofold. First, in equilibrium or static quantum phases, twisted boundary conditions are typically implemented in ground states or partition functions. However, in non-equilibrium and dynamical settings, these objects are no longer applicable, requiring us to seek an appropriate alternative. Here, we consider the spectral form factor (SFF)
of unitary time evolution operators 
as a key diagnostic and examine the effect of twisted boundary conditions in this context.
Second, we generalize twisted boundary conditions to discrete time-translation symmetry, contrasting with the more conventional approach of twisting via a conserved on-site symmetry, such as the U(1) symmetry associated with conserved particle number. This extension offers new insights into the characterization of dynamical phases beyond equilibrium frameworks.
Specifically, we build upon our previous works 
\cite{PhysRevB.106.155128,doi:10.7566/JPSCP.38.011175},
where we introduced the energy-twisted boundary condition 
\footnote{
While the systems studied in Refs.\ \cite{PhysRevB.106.155128,doi:10.7566/JPSCP.38.011175}
conserve energy—i.e., they possess continuous time-translation symmetry—the Floquet systems considered in this work do not. 
Thus, we distinguish them by using different terminology
\label{footnote:twist}}
which applies twisted boundary conditions using time-translation symmetry in an equilibrium setting, to investigate thermal transport properties of (1+1)-dimensional many-body quantum systems. For related studies, see also 
Refs.\ \cite{PhysRevResearch.3.023118, PhysRevB.108.075108,
Kishony2025}.

To place our exploration in the proper context, let us begin by reviewing the concept of twisted boundary conditions in the study of static phases of matter with particle number conservation.
In static electronic systems, the U(1) twisted boundary condition 
\begin{align}
    \psi(x+L)=e^{i\phi}\psi(x)
\end{align} 
of the electronic field $\psi$ on a ring of circumference $L$ has been used to characterize electronic phases.
Specifically, the Anderson localization is diagnosed by the sensitivity against the U(1) twisted boundary condition \cite{JTEdwards_1972}.
Spatially localized electronic states have exponentially small coherence between two points much further apart than the localization length,
and are less sensitive to the boundary condition 
for large system sizes.
The sensitivity is measured by the ratio between the mean level spacing and the Thouless energy $E_\text{Th}$, which is the difference of each energy level between periodic and antiperiodic boundary conditions.
Typically, $E_\text{Th}=\hbar D/L^2$ when the diffusion constant is $D$.

Another example utilizing the U(1) twisted boundary condition is in superfluidity (superconductivity).
Superfluid is characterized by spontaneously broken U(1) symmetry.
In the bulk, electrons form Cooper pairs, behave collectively, and share a coherent U(1) phase.
Imposing the U(1) twisted boundary condition interferes with this coherence destructively.
Specifically, one can calculate the stiffness of superfluid (superconductors)
\begin{align}
    D_\text{Meissner}
    =
    \frac{2}{L}\frac{d^2F}{d\phi^2}\bigg|_{\phi=0},
    \label{eq:meissner}
\end{align}
which is known as the Meissner stiffness \cite{PhysRevB.47.7995,PhysRevB.51.10915,Resta_2018}, where $F$ is the free energy.
The finiteness of $D_\text{Meissner}$ is a criterion for superconductivity.
Experimentally, a similar sensitivity appears as the Little-Parks effect \cite{tinkham2004introduction}.
Consider a thin superconductor in a cylinder geometry.
The U(1) twisted boundary condition is encoded as the magnetic flux $\phi \hbar/e$ threading into the hole of the cylinder.
The magnetic flux induces a circular supercurrent and increases the kinetic energy, resulting in the decrease of the superconducting critical temperature.
After threading half of the quantum flux $h/2e$, the coherent Cooper pairs experience no flux.
The critical temperature thus oscillates periodically as a function of the magnetic flux.

A prominent example of the spontaneously symmetry broken phase unique to dynamical systems is time crystals, where time-translation symmetry is broken 
in essentially the same way as (spatial) crystals break space-translation symmetry \cite{PhysRevLett.109.160401}.
In particular, in periodically driven systems, time crystals that oscillate in a different period from the driving one is referred to as discrete time crystals (DTCs) \cite{PhysRevA.91.033617, PhysRevLett.116.250401, PhysRevLett.117.090402, PhysRevLett.118.030401,*PhysRevLett.118.269901}.
To stabilize the DTC phase, one needs 
some mechanisms such as many-body localization, which is a dynamical analogue of Anderson localization, and prethermalization.
Otherwise, a system is driven into an infinite temperature which is irrelevant to the initial state and featureless \cite{PhysRevX.4.041048, PhysRevE.90.012110, PONTE2015196}.
We focus on the many-body localized case in this paper.

In this study, we impose a boundary condition to diagnose the chaotic and DTC phases in the same spirit as the U(1) twisted boundary condition for the Anderson localization and superconductivity.
From the analogy to superconductors, one can expect that dynamical phases breaking time-translation symmetry could be diagnosed by a boundary condition twisted by time translation.
We refer to it as
the time-translation twist
which is the discrete time-translation symmetric version of the energy-twisted boundary condition
\cite{PhysRevB.106.155128,doi:10.7566/JPSCP.38.011175}.
Schematically, it is defined by
\begin{align}
    \psi(x+L,t)=\psi(x,t+a).
    \label{eq:energytwistedboundarycondition}
\end{align}
With this boundary condition, when a particle goes around the spatial circle, time is translated by $a$.
See Sec.\ \ref{sec IIb} for a more precise definition.

Our reasoning for using the twisted boundary condition, based on time translation symmetry, as a diagnostic tool for detecting DTC phases is as follows. In a DTC, the entire system oscillates coherently with a period that is an integer multiple of the driving period. This coherent oscillation can be disrupted by the time-translation twist.
Specifically, when one side of the boundary is translated in the temporal direction, the oscillation phase becomes mismatched across the boundary, leading to destructive interference.
In a DTC, Floquet states are invariant under discrete time translation by an integer multiple of the driving period, whereas in other phases, they follow the driving period itself. 
The phase mismatch introduced by the time-translation twist
inhibits the formation of the time crystal. 
However, when twisting 
the period of the DTC's oscillation, 
the oscillation phase realigns 
and the coherent oscillation is restored.
As a result, DTCs exhibit a periodic response to the time-translation twist
as a function of 
$a$ in Eq.\   \eqref{eq:energytwistedboundarycondition}.
This phenomenon can be seen as a DTC analogue of the Little-Parks effect,
in the sense that symmetry-broken phases are diagnosed by boundary conditions twisted by the broken symmetry.
Notice that the Little-Parks effect in superconductors is periodic in units of the magnetic flux quantum, while the analogous effect in DTCs is periodic in units of the system's oscillation period.

Quantum chaotic phases are distinguished from many-body localized phases (including the DTC phase), e.g., by the statistics of (quasi) energy spectrum.
The chaotic phases show universal spectral correlation that obeys the random matrix theory \cite{Meh2004}.
In addition to quantum many-body systems, the chaotic dynamics has also been studied in random unitary circuits \cite{annurev:/content/journals/10.1146/annurev-conmatphys-031720-030658}.
In this work, we study the chaotic and DTC phases using the disordered kicked Ising model.

This paper is organized as follows.
In Sec.~\ref{sec:preliminaries}, we review the spectral form factor (SFF) and introduce a few kinds of boundary conditions that we use in this paper. We also show how to implement them in Floquet systems as tensor networks.
In Sec.~\ref{sec:kickedisingmodel}, we impose the time-translation twist
to the disordered kicked Ising model and show that the chaotic and DTC phases are characterized in terms of either short- or long-time behavior of the SFF.
Finally, we conclude in Sec.~\ref{sec:conclusion}.

\section{Preliminaries}\label{sec:preliminaries}

\subsection{Spectral form factor}

In this work, we are interested in the time-evolution operator of many-body quantum systems. 
First, we are interested in Floquet time evolution operators.
When a periodically time-dependent Hamiltonian $H(t+T)=H(t)$ with period $T$ is given, the Floquet dynamics is characterized by the Floquet operator
\begin{align}
    U_F =
    \mathcal{T}\exp\left[-i\int_0^T dt H(t)\right],
    \label{UF}
\end{align}
where $\mathcal{T}$ is the time-ordering operator.

The object of our main focus is the spectral form factor (SFF), 
which is the Fourier transform of the spectral correlator, defined by
\begin{align}
    K(t)=|\text{Tr}\, U_F^t|^2
    =
    \sum_{j,k}e^{i(\theta_j-\theta_k)t},
    \label{eq:sff}
\end{align}
where the discrete time $t$ 
represents
integer multiples of the period $T$
and $\theta_j\in[0,2\pi/T)$ is a quasienergy of the Floquet operator $U_F$.
The SFF can be characterized by multiple time scales. 
For example, the Thouless time represents the time scale at which the random-matrix-theory behavior sets in. 
Another example is the Heisenberg time at which the SFF saturates.

In chaotic systems, the SFF is dependent on microscopic details before the Thouless time, 
grows in the same manner as the random matrix theory,
and then saturates at the Heisenberg time as 
$K(t)\sim q^L$ where $q$ is the local Hilbert space dimension and $L$ is the system size \cite{PhysRevLett.121.060601, PhysRevLett.121.264101, PhysRevX.8.021062, PhysRevX.8.041019, PhysRevResearch.2.043403}.
The SFF of the random matrix theory with time-reversal symmetry obeys the circular orthogonal ensemble (COE) while that without it obeys the circular unitary ensemble (CUE) given, respectively, by
\begin{align}
    K(t)
    =
    \left\{
    \begin{array}{ll}
        2t-t\ln(1+2t/q^L)\quad &(\text{COE})\\
        t\quad &(\text{CUE})
    \end{array}
    \right..
    \label{eq:spectralformfactor_randommatrixtheory}
\end{align}
On the other hand, in many-body localized systems, the spectrum obeys the Poisson distribution and the SFF is saturated at $K(t)=q^L$ from the beginning \cite{PhysRevE.102.062144, PhysRevResearch.3.L012019}.

\subsection{Time-translation twist in Floquet systems}
\label{sec IIb}

\begin{figure}
\centering
\includegraphics[width=0.5\textwidth]{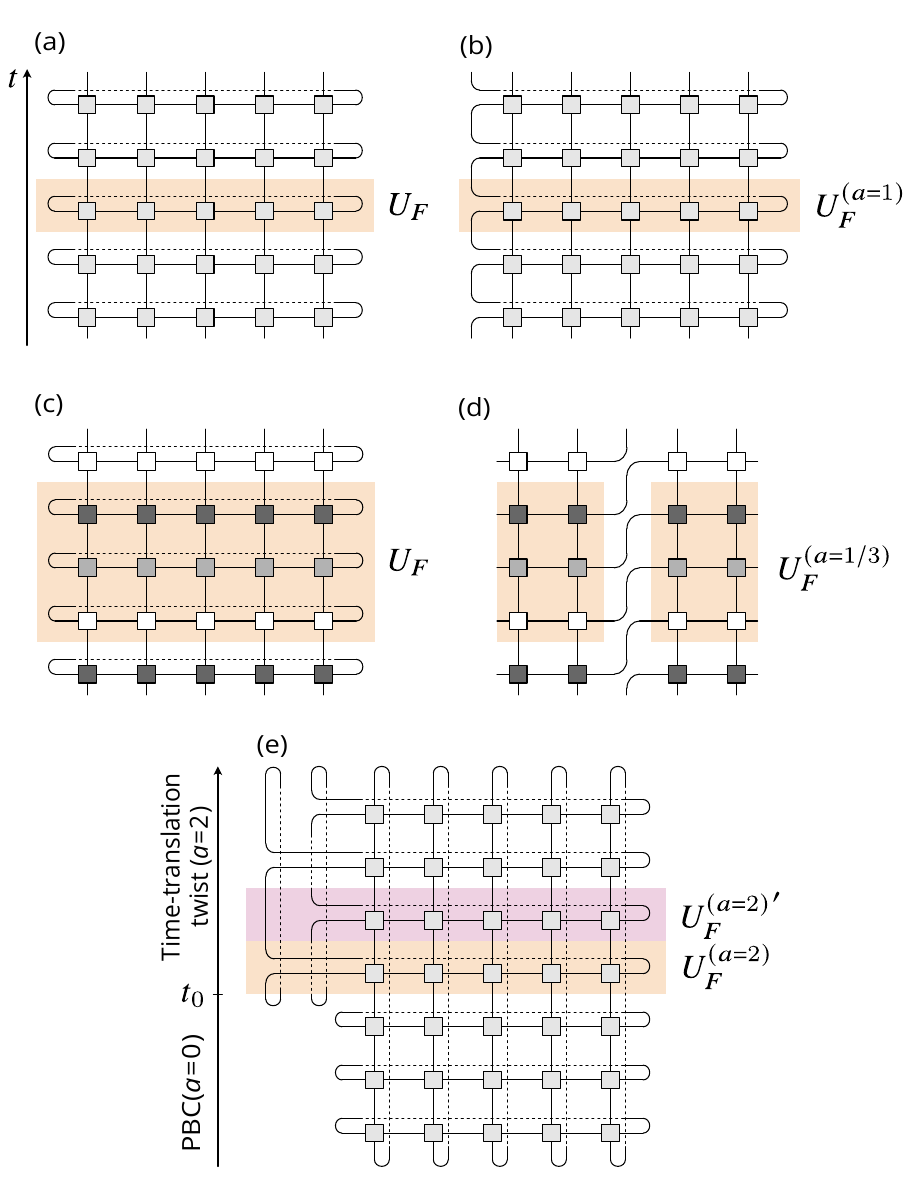}
\caption{
    Tensor-network representation of the time-translation twist.
    (a-b) The product of Floquet operators by the ordinary temporal transfer matrix and that of twisted Floquet operators with $a=1$, respectively.
    (c-d) The product of fractionalized Floquet operators and of twisted Floquet operators with $a=1/3$, respectively.
    A consecutive set of rows highlighted by orange rectangles is the unit of the temporal transfer matrix.
    Black, gray, and white squares in (c) and (d) represent different roles of operators in a single period.
    (e) The boundary condition is switched from the periodic one to the one with time-translation twist
    with $a=2$ at $t_0$.
    In this figure, the trace that connects the top and bottom legs is included.
    \label{fig:tensornetwork}
}
\end{figure}

Floquet operators and the corresponding SFF on a one-dimensional 
lattice ($j\in[1,L]$) can be represented as 
tensor networks via the matrix product operator \cite{Pirvu_2010} as in Fig.~\ref{fig:tensornetwork}.
The horizontal and vertical directions are spatial and temporal, respectively.
Each row corresponds to a single period of the Floquet operator at $t$ 
which is an integer multiple of the driving period $T$.
In the following, we set $T=1$ for brevity.
Under the periodic boundary condition, the rightmost operator is connected to the leftmost one on the same row [Fig.~\ref{fig:tensornetwork}(a)].

Upon application of the time-translation twist, the horizontal links between the rightmost and leftmost operators are cut and reconnected between different rows [Fig.~\ref{fig:tensornetwork}(b)].
Specifically, operators at $(j=L,t)$ are connected to $(j=1,t+a)$.
This method can be applied to any integer $a=n$. 
Notice that the resulting Floquet operator $U_F^{(a=n)}$ has $n$ extra legs that traverse the boundary [Fig.~\ref{fig:tensornetwork}(b)], and the dimension of $U_F^{(a=n)}$ is $2^n$ times larger than the untwisted one.
When $n\ge 2$, the twisted Floquet operators have $n$ distinct forms that alternate [see pink and orange operators in Fig.~\ref{fig:tensornetwork}(e)].
In addition to this basic time-translation twist, we extend it in two
ways as follows to capture characteristics of the dynamical phases.

The first extension is the fractionalization of the twist as $a=p/q$ where $p,q\in\mathbb{Z}$.
Specifically, we decompose the Floquet operator into $q$ parts as
\begin{align}
    U_F = U_{(q-1)F}\cdots U_{1F}U_{0F},
    \label{eq:floquetunitary_decomposition}
\end{align} 
where $U_{rF}$ is the time-evolution operator between $t\in[r/q,(r+1)/q]$ [Fig.~\ref{fig:tensornetwork}(c)].
Given a time-periodic Hamiltonian, this decomposition is unique as we simply decompose the evolution time into $1/q$.
The fractional time-translation twist
is then imposed in a similar fashion as the integer case, that is, by cutting the horizontal links and reconnecting operators at $p$th neighboring rows [Fig.~\ref{fig:tensornetwork}(d)].
We refer to this twist as the fractional twist, which will be applied in the study of the short-time behavior in Sec.~\ref{sec:shorttime}.

The second extension is to change the boundary condition at some time $t_0$ during the Floquet evolution [Fig.~\ref{fig:tensornetwork}(e)].
Before $t_0$, the system evolves under the periodic boundary condition, and then the boundary condition is switched to the one with the time-translation twist.
Specifically, for $a=1$, the time evolution operator is given by
\begin{align}
    U(t)=
    \left\{
    \begin{array}{ll}
        U_{F}^t \quad &(t\le t_0)\\[+3pt]
        \left[{U_{F}^{(a=1)}}\right]^{t-t_0}(\sigma^0\otimes U_{F}^{t_0}) \quad &(t>t_0)
    \end{array}
    \right.,
\end{align}
where $\sigma^0$ is the $2\times 2$ identity matrix.
We doubled the dimension of the Hilbert space at $t_0$ to accommodate the twisted Floquet operator that has one extra leg.
Equivalently, we can add an auxiliary qubit at the beginning and then evolve it by $\sigma^0$ (that is, by a zero-matrix Hamiltonian) until $t_0$.  
Similarly, for $a=2$, the time-evolution operator reads $U(t)=U_{F}^t$ until $t_0$, and then two types of the Floquet operators alternate as [Fig.~\ref{fig:tensornetwork}(e)]
\begin{align}
    U(t)=
    \left\{
    \begin{array}{ll}
        {U_{F}^{(a=2)}}'U(t-1) \quad &(t-t_0\text{: even})\\
        U_{F}^{(a=2)}U(t-1) \quad &(t-t_0\text{: odd})
    \end{array}
    \right.,
\end{align}
where $U(t_0)=\sigma_0\otimes\sigma_0\otimes U_{F}^{t_0}$.
We will apply this boundary condition in the study of the long-time behavior in Sec.~\ref{eq:long-time}.

\section{Kicked Ising model}
\label{sec:kickedisingmodel}

In this section, we consider the DTC
realized in the one-dimensional kicked Ising model under the time-translation twist.

Let us begin by reviewing the definition and the phase diagram of the kicked Ising model in one dimension 
following \cite{PhysRevLett.117.090402,PhysRevLett.118.030401,*PhysRevLett.118.269901,PhysRevLett.121.264101}.
The Hamiltonian is defined by
\begin{align}
    H_{KI}[\bm{J},\bm{h};t]=H_I[\bm{J},\bm{h}]+\delta_p(t)H_K,
    \label{eq:kickedisinghamiltonian}
\end{align}
where $\delta_p(t)=\sum_{n\in\mathbb{Z}}\delta(t-n)$ specifies the periodic kick timing with period $ T=1$, and
\begin{align}
    &H_I[\bm{J},\bm{h}]
    =
    \sum_{j}\left(J_j\sigma_j^z\sigma_{j+1}^z+h_j\sigma_j^z\right),
    \label{eq:kickedisinghamiltonian_ising} \\
    &H_K 
    =
    b\sum_{j}\sigma_j^x.
    \label{eq:kickedisinghamiltonian_kick}
\end{align}
Here, $J_j$ and $h_j$ are taken to be random variables
with mean values ${\bar J}$ and $\bar{h}$ 
and variances $\sigma_J$ and $\sigma_h$, 
respectively.
Unless stated otherwise, we will consider uniform distributions.
The corresponding Floquet operator is 
\begin{align}
    U_{KI}=e^{-iH_K}e^{-iH_I[\bm{J},\bm{h}]}.
    \label{eq:kickedising_floquetoperator}
\end{align}
In each period, the evolution with the spatially disordered Ising coupling and longitudinal field is followed by the instantaneous application of the transverse field (kick).

In the absence of the longitudinal field $h_j=0$, the kicked Ising model is mapped to a free Majorana fermion model by the Jordan-Wigner transformation and is exactly solvable.
The free fermion model has 4 phases, each of which is characterized by the presence/absence of the boundary Majorana modes at quasienergies $0$ and/or $\pi$ \cite{PhysRevLett.106.220402, PhysRevA.86.063627, PhysRevB.88.155133} [see Fig.~\ref{fig:sff_boundarychange}(a)].
These phases 
continue to exist even when we turn on 
the longitudinal field \cite{PhysRevLett.116.250401, PhysRevB.93.245145, PhysRevB.93.245146}.
Following \cite{PhysRevLett.116.250401}, we call these phases as PM, 0SG, $\pi$SG, and 0$\pi$PM phases, respectively.
The center of the phase diagram $(b,\bar{J})=(\pi/4,\pi/4)$ is self-dual 
under the transformation 
interchanging space and time \cite{Akila_2016} and is chaotic \cite{PhysRevLett.121.264101}.
The $\pi$SG phase shows the DTC, in which all spins oscillate coherently at twice the driving period \cite{PhysRevLett.116.250401, PhysRevLett.117.090402, PhysRevLett.118.030401,*PhysRevLett.118.269901}.
This phase is the focus of this section.
The 0$\pi$PM phase shows a boundary time crystal \cite{PhysRevB.93.245145,khemani2019briefhistorytimecrystals}, in which the boundary spins oscillate at twice the driving period under the open boundary condition.

The SFF of the kicked Ising model in the many-body localized and chaotic phases has been studied, e.g. in \cite{PhysRevLett.121.060601, PhysRevLett.121.264101, PhysRevX.8.021062, PhysRevX.8.041019, PhysRevResearch.2.043403, PhysRevLett.121.264101, PhysRevResearch.3.023118}.
We will incorporate the time-translation twist
into these studies.

We will discuss the two limiting regimes separately: the short-time limit ($L > t$) and the long-time limit ($L < t$), where $L$ is the length of the one-dimensional chain and $t$ is the total number of time steps.
From a technical perspective, due to the unsolvability of the model, the linear dimensions in either the spatial or temporal direction are constrained, requiring separate analyses of these two limits.

More importantly, these limits capture different physical behaviors. 
In the short-time regime, as we will show below, the vanishing of the SFF under a half-period twist is a consequence of the DTC. In the long-time regime, twisting the boundary condition after the formation of a DTC reduces the SFF, with the effect depending on whether the twist period is even or odd. In both cases, the imposed twists introduce phase mismatches, either in the driving or the oscillation, across the boundary.

It is important to note that we adopt different definitions of DTC in these two cases. 
The long-time DTC corresponds to the definition used in previous studies [Fig.\ \ref{fig:sff_boundarychange}(a)], the short-time DTC is defined via the spectral Lyapunov exponent
\cite{PhysRevResearch.3.023118}. 
Due to this distinction, the phase diagrams differ between the two cases.
Notably, the short-time phase diagram depends on the parameter $\bar{h}$, while the long-time phase diagram does not.
We distinguish regions in the phase diagrams of the short-time behavior by the term "region", 
since the definition and the detection method shown in Sec.~\ref{sec IIb} are not relevant to the rigidity of a "phase" as will be defined by the long-time behavior.

\subsection{Short-time behavior}\label{sec:shorttime}

Let us first study the short-time behavior of the SFF with time-translation twist. 
In this regime (limit), we can utilize the duality of the kicked Ising model exchanging space and time \cite{Akila_2016}.
Using this duality, the row-to-row transfer matrix (\ref{eq:kickedising_floquetoperator}) is transformed to the column-to-column transfer matrix satisfying $\text{Tr}\,U_{KI}^t=\text{Tr}'\,\tilde{U}_{KI}^{(L)}\cdots \tilde{U}_{KI}^{(1)}$,
where $\text{Tr}'$ is taken over the 
dual Hilbert space $(\mathbb{C}^2)^{\otimes t}$, on which the temporal (virtual) spin operator $\tau_k\,(k\in[1,t])$ acts.
In the tensor network representation, each real spin degrees of freedom corresponds to a vertical leg while each temporal one to a horizontal leg.
Specifically,
\begin{align}
    \tilde{U}_{KI}^{(j)}
    =
    c^te^{-i\tilde{H}_K[J_j]}e^{-i\tilde{H}_I[h_j]},
    \label{eq:ctctransfermatrix}
\end{align}
where $c=(-\sin 2b \sin 2J)^{1/2}$ and by using the Pauli matrices $\tau_k$
acting on the Hilbert space of spins along the temporal direction $k\in[1,t]$
\begin{align}
    &\tilde{H}_I[h_j]
    =
    \sum_{k=1}^t \left(b^\ast \tau_k^z\tau_k^z +h_j\tau_k^z\right),\\
    &\tilde{H}_K[J_j]
    =
    J_j^\ast\sum_{k=1}^t \tau_k^x.
\end{align}
Here, the dual coefficients are defined by
\begin{align}
    A^\ast
    =
    -\frac{\pi}{4}
    -
    \frac{i}{2}\log\tan A.
\end{align}
With the dual picture, the column-to-column transfer matrix is encoded on a tensor network.
Even with the fractionally decomposed Floquet operators (\ref{eq:floquetunitary_decomposition}), a similar procedure leads to the tensor network representation.
The dual picture is useful in the evaluation of the time-translation twist
since the transfer matrix is invariant under time translation.
The time-translation twist 
is implemented by a translation along the time direction
at the end of the product of the column-to-column transfer matrix.
Fortunately, the numerical cost of evaluating the SFF is unchanged even with a fractional twist, since the most parts of the decomposed Floquet operators in Eq.~(\ref{eq:floquetunitary_decomposition}) consist only of $\sigma^z$.

Notice that after a fractional twist, the timing of the kick is not smoothly connected across the boundary [Fig.~\ref{fig:tensornetwork}(d)].
In a DTC regime, the sudden change in kick timing at the boundary cannot extend to the bulk so that the kick timing difference is smoothed out in the thermodynamic limit as discrete time-translation symmetry is broken.
This is analogous to superconductors, where the U(1) twisted boundary condition cannot be transformed to a uniform vector potential due to the breaking of U(1) gauge symmetry.

We numerically evaluated the ratio $\langle K(t,a)\rangle/\langle K(t,0)\rangle$ between the twisted and untwisted averaged SFFs.
\begin{figure}
    \centering
    \includegraphics[width=0.49\textwidth]{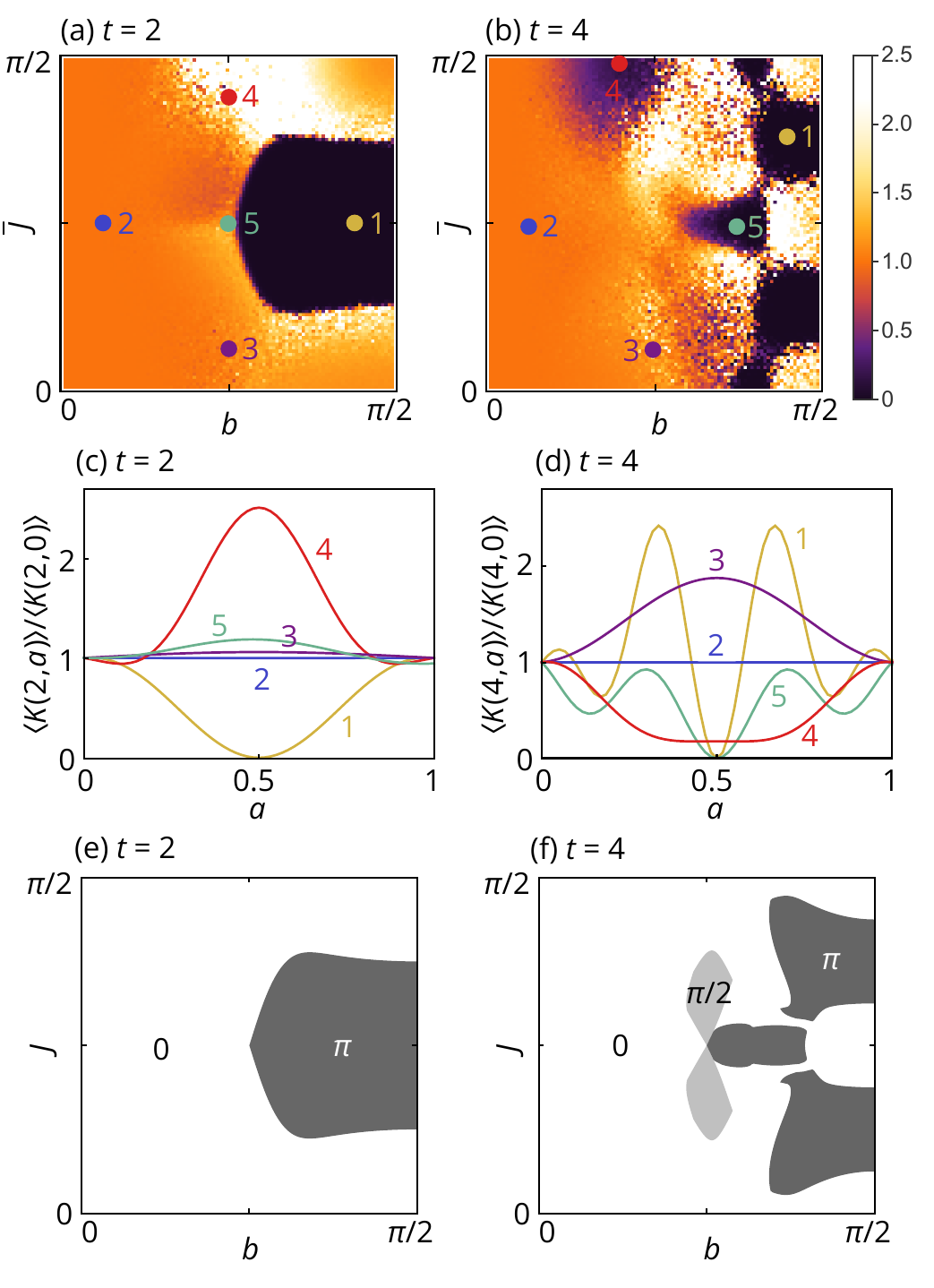}
    \caption{(a-b) The ratio of half-twisted and untwisted SFFs with $L=30$ is shown for $t=2$ and $t=4$, respectively.
        The number of disorder configurations is 10000 for (a) and 1000 for (b).
        (c-d) The SFF as a function of twist for some points indicated in (a) and (b).
        The frequency of the largest eigenvalue state of the bulk column-to-column transfer matrix for (e) $t=2$ and (f) $t=4$.
    \label{fig:kickedising_sffhalf}
    }
\end{figure}
In particular, we choose $t=2$ and $4$ since the SFF vanishes for small odd $t$.
The SFF was averaged over a uniformly distributed longitudinal field and Ising coupling 
with mean $\bar{h}=0.6$ and variance $\sigma_h=\sigma_J=\pi/10$.
Figs.~\ref{fig:kickedising_sffhalf} (c) and (d)
show the ratio as a function of the twisted period $a$ for representative points in the phase diagram,
while Figs.~\ref{fig:kickedising_sffhalf} (a) and (b)
present the ratio at a half-period twist $a=1/2$. 
In the dark region of Figs.~\ref{fig:kickedising_sffhalf} (a) and (b), the ratio vanishes at a half-period twist.
As we will see momentarily, the vanishing SFF at $a=1/2$ is 
a signature of the DTC.

We define the DTC 
in the short-time limit as follows. 
The column-to-column transfer matrix is invariant under time translation by integer times the driving period.
Thus, it can be block-diagonalized by the frequency $\omega=2\pi n/t \,(n\in[0,t-1])$ (notice that we are working on the Hilbert space along the temporal direction, and we set the driving period $T=1$).
The product of the column-to-column transfer matrix grows exponentially by the leading Lyapunov exponent, which is the largest one in the respective frequency subspace.
We classify regimes in the short-time behavior based on the frequency associated with the leading Lyapunov exponent, since its oscillation is dominant in sufficiently large systems.
In particular, the DTC of our focus corresponds to $\omega=\pi$.

With this definition, we can directly link the vanishing of the SFF at $a=1/2$
to the DTC order.
For $t=2$, we can show that the SFF at $a=1/2$ vanishes when the leading Lyapunov exponent is from $\omega=\pi$ subspace (see Appendix \ref{app:t2} for details).
The frequency $\omega$ of a state is defined by the eigenvalue $e^{i\omega}$ of the time translation operator 
$\hat{T}$, which is defined by 
$\hat{T}|\tau_1,\tau_2\rangle=|\tau_2,\tau_1\rangle$ where $\tau_k^z|\tau_1,\tau_2\rangle=\tau_k|\tau_1,\tau_2\rangle$.
The Hilbert space is spanned by four states, one of which, $|\pi\rangle$, has frequency $\pi$,
and the remaining three, $|0_j\rangle$, have 0.
Specifically,
\begin{align}
    &|\pi\rangle=|\!\uparrow\downarrow\rangle-|\!\downarrow\uparrow\rangle, \\
    &|0_1\rangle = |\!\uparrow\uparrow\rangle,\quad
    |0_2\rangle = |\!\downarrow\downarrow\rangle,\quad
    |0_3\rangle = |\!\uparrow\downarrow\rangle+|\!\downarrow\uparrow\rangle.
\end{align}
Since the $\omega=\pi$ subspace has a single state, it must be an eigenstate of the column-to-column transfer matrix regardless of the random couplings $J_j$ and $h_j$.
As for the column-to-column transfer matrix across the boundary, the eigenvalue of $|\pi\rangle$ becomes 0 when $a=1/2$.
As a result, the DTC regime characterized by $\omega=\pi$ must accompany a vanishing SFF 
at $a=1/2$.
The shaded region in Fig.~\ref{fig:kickedising_sffhalf}(e) is the DTC regime determined by the frequency of the leading Lyapunov exponent without disorder ($\sigma_J=\sigma_h=0$), which agrees with the dark region in Fig.~\ref{fig:kickedising_sffhalf}(a).
A similar line of argument can be applied to $t=4$, though the details are 
more complicated (see Appendix \ref{app:t4}).

Although we previously fixed $\bar{h} = 0.6$, repeating the analysis for different values of $\bar{h}$ reveals that the resulting phase 
diagrams—counterparts of  Figs.~\ref{fig:kickedising_sffhalf}(e) and (f)- do depend on $\bar{h}$ except for $b=\pi/2$.
We also note that, other than the DTC regime, the response against the time-translation twist
is not symmetric with respect to $\bar{J}=\pi/4$ -- this should be contrasted with the long-time behavior, as we will see later.
Meanwhile, the 0SG regime is almost insensitive to the time-translation twist.

As $t$ increases, the vanishing of the SFF at a half twist could be less clear unless we consider the thermodynamic limit [see discussion around (\ref{eq:ratiovanish_app}) in Appendix \ref{sec:bpi2}].
On the other hand, the phase diagram determined by the frequency of the leading Lyapunov exponent is divided into a number of small regions since the number of possible frequency increases as $t$ increases.
Taking these into account, the coincidence of the vanishing SFF region as in Figs.~\ref{fig:kickedising_sffhalf}(a) and (b) and the $\pi$ dominant region as in Figs.~\ref{fig:kickedising_sffhalf}(e) and (f) becomes less clear for large $t$.
The same applies to smaller system sizes.

\subsection{Long-time behavior}
\label{eq:long-time}

We now discuss the long-time behaviors.
In the long-time regimes, the row-to-row transfer matrix is more convenient than the column-to-column transfer matrix.
The product of the row-to-row transfer matrix evolves along the temporal direction.
We incorporate the time-translation twist
by using the matrix product operator \cite{Pirvu_2010}.
The matrix element of the row-to-row transfer matrix is written as
\begin{align}
    \langle \{s'\}|U_{KI}|\{s\}\rangle
    =
    \text{tr}\, A^{(L)}_{s_L's_L}\cdots A^{(1)}_{s_1's_1},
    \label{eq:kickedising_temporaltransfermatrix_component}
\end{align}
where $\{s\}=\{s_1,s_2,\cdots,s_L\}$, $\{s'\}=\{s_1',s_2',\cdots,s_L'\}$ are legs going downward and upward, respectively, and
$A_{s's}$ is a $2\times 2$ matrix acting on an auxiliary spin space denoted by a horizontal leg across the boundary [Fig.~\ref{fig:tensornetwork}(a)].
The trace is taken over the auxiliary space.
We decompose the Floquet operator into the Ising-coupling part and the other part as $U_{KI}=U_1[b,\bm{h}]U_0[\bm{J}]$.
The Ising-coupling part is written as 
\begin{align}
    &\langle \{s'\}|U_0[\bm{J}]|\{s\}\rangle 
    =
    \left(\prod_j \delta_{s_j's_j}\right)
    \text{tr}\, \tilde{A}^{(L)}_{s_L}\cdots \tilde{A}^{(1)}_{s_1},
\end{align}
where
\begin{align}
    \tilde{A}^{(j)}_{s_j}&=
    \begin{pmatrix}
        \sqrt{\cos J_j\cos J_{j-1}} & -is_j\sqrt{\cos J_j\sin J_{j-1}} \\
        s_j\sqrt{\sin J_j\cos J_{j-1}} & -i\sqrt{\sin J_j\sin J_{j-1}}
    \end{pmatrix}.
\end{align}
Then we obtain
\begin{align}
    A^{(j)}_{s_j's_j}
    =
    \left(e^{-ib\sigma_j^x}e^{-ih_j\sigma_j^z}\right)_{s_j's_j}
    \tilde{A}^{(j)}_{s_j}.
\end{align}

Twisting the boundary condition is incorporated by cutting and reconnecting the horizontal legs across the boundary [Fig.~\ref{fig:tensornetwork}(b)].
There are $a$ additional legs in the row-to-row transfer matrix when we twist the boundary condition by integer $a$.
The total Hilbert space dimension $2^{L+a}$ grows exponentially as $a$ is increased.
A fractional twist $a=p/q$ is implemented in the same way as the integer twist via Eq.~(\ref{eq:floquetunitary_decomposition}).

First, we confirmed that the long-time behavior of the DTC is not disturbed by the twist, if we impose it from the outset, that is, if the preparation time 
in Fig.~\ref{fig:tensornetwork}(e) is set to zero, $t_0=0$
(see Appendix \ref{app:t00} for details). 
This is because another form of the DTC, distinct from the one realized under the periodic boundary condition, that has a spatially non-uniform oscillation phase, is realized under the presence of the twist.
So, we need to wait until a dynamical phase develops under the periodic boundary condition 
(by taking $t_0>0$) 
and then disturb the phase by imposing the time-translation twist.
This setup is analogous
to the Little-Parks effect, where the U(1) twisted boundary condition is imposed after superconductivity has been established.

We numerically evaluate the averaged SFF of the kicked Ising model of length $L=8$ with the same parameters as before.
\begin{figure}
    \centering
    \includegraphics[width=0.48\textwidth]{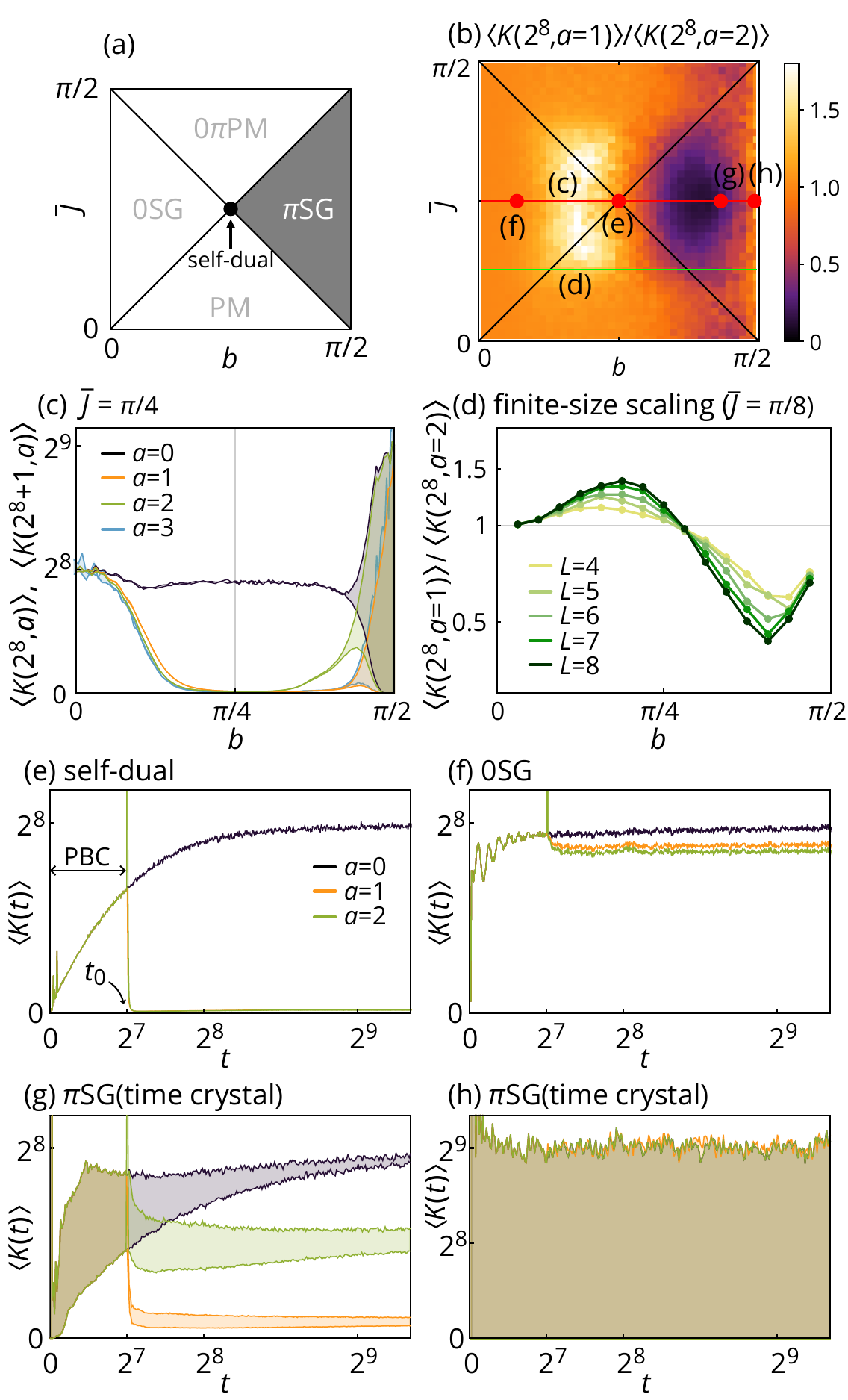}
    \caption{
        (a) The phase diagram of the kicked Ising model. Notice that when $h_j=0$, the phase diagram is invariant under $b\to b+\pi$, $\bar{J}\to \bar{J}+\pi$, $b\to -b$ and $\bar{J}\to -\bar{J}$ due to the corresponding symmetry of the Floquet operator (\ref{eq:kickedising_floquetoperator}).
        (b) The phase diagram showing the ratio of the averaged SFFs at $t=2^8$ 
        for $a=1$ and $2$,
        where the preparation time is $t_0=2^7$.
        (c) The averaged SFF at $t=2^8$ and $t=2^8+1$ as a function of $b$ at $\bar{J}=\pi/4$ [red line in (b)].
        (d) The finite-size scaling of the SFF ratio from $L=4$ to $8$ at $\bar{J}=\pi/8$ [green line in (b)].
        (e)-(h), the averaged SFF of $L=8$ when the boundary condition is switched at $t_0=2^7$ is shown for $a=0,1$, and $2$.
        Each figure corresponds to the parameter given in (b).
        We averaged over 1000 samples for (b) and $a=3$ in (c), while 10000 samples otherwise.
        \label{fig:sff_boundarychange}
    }
\end{figure}
We employ the preparation time $t_0=2^7$ since the oscillation of the SFF in the DTC phase is well-developed but still not weakened at this time as can be seen from the black curve in Fig.~\ref{fig:sff_boundarychange}(g)
\footnote{We confirmed that the conclusions about the long-time behavior are not changed even if we choose either $t_0=2,4,\cdots,$ or $2^8$ as the preparation time.}.
Figs.~\ref{fig:sff_boundarychange}(e)--(h) are the SFF for 
as a function of time $t$ at some points indicated in Fig.~\ref{fig:sff_boundarychange}(b).
In each figure, three lines correspond to $a=0$, $1$, and $2$ respectively.

The effect of the time-translation twist
on the DTC is most pronounced in 
Fig.~\ref{fig:sff_boundarychange}(g),
which shows the averaged SFF deep within the DTC phase.
While twists with $a=1$ and $a=2$ both reduce the SFF, 
the twist with $a=1$ disturbs the DTC order more significantly than that with $a=2$.
This is because the oscillation phase of the DTC (mis)matches across the boundary when ${a=2}$(${a=1}$).
In this way, the time-translation twist
serves as a probe for detecting the DTC order.
On the other hand, within the 0SG phase, the SFF at late times is less sensitive to the twist, and there is no significant difference 
between the cases with $a=1$ and $a=2$.
In Fig.~\ref{fig:sff_boundarychange}(b),
we plot the ratio $\langle K(2^8,a=1)\rangle/\langle K(2^8,a=2)\rangle$,
which deviates from unity most prominently in the DTC phase.
In Fig.~\ref{fig:sff_boundarychange}(c),
we also plot the SFF at $t=2^8$ and $t=2^8+1$ for $\bar{J}=\pi/4$ [the red line in Fig.~\ref{fig:sff_boundarychange}(b)] as a function of $b$.
For reference, we also plot the $a=3$ case at $t=224$ and $t=225$, which almost agrees with $a=1$.
This implies the distinction between even and odd $a$.
In Fig.~\ref{fig:sff_boundarychange}(d), we plot the finite-size scaling of the ratio along $\bar{J}=\pi/8$ [the green line in Fig.~\ref{fig:sff_boundarychange}(b)] from $L=4$ to $L=8$, which shows a phase transition at $b\simeq0.28\pi$.
Here, we used a common preparation time $t_0=2^7$ and evaluated the ratio at $t=2^8$ for all the system sizes.

We note that at microscopic time scales, DTC behavior can be observed through oscillations in the SFF. However, these oscillations generally decay at late times. In contrast, sensitivity to the time-translation twist
remains a robust probe of DTC order, even after temporal oscillations have faded. The line $b = \pi/2$ is somewhat special in this regard, as DTC oscillations persist to later times, potentially “masking” the changes induced by the time-translation twist
[Fig.~\ref{fig:sff_boundarychange}(h)].
In Appendix \ref{sec:bpi2}, we analytically confirm two features of the SFF at $b=\pi/2$:
(i) the long-time behavior of the SFF is the oscillation between 0 and $2^{L+1}$ [Fig.~\ref{fig:sff_boundarychange}(h)];
and (ii) the ratio between untwisted and a half-period twisted SFF vanishes 
in the short time regime (Fig.~\ref{fig:kickedising_sffhalf}).

Finally, while our main focus has been on the long-time behavior of the SFF under the time-translation twist
as a probe of DTC order, let us briefly comment on the SFF at shorter times, particularly near $t=t_0$.
First, we note that the sharp peak 
in the SFF around $t_0=2^7$ in Fig.~\ref{fig:sff_boundarychange}(e)--(g) may 
arise from the sudden change of the Hilbert space size.
Additionally, in general, the SFF exhibits a sharp drop 
right after $t_0$.
Let the Floquet operator before and after $t_0$ be denoted by $U_F$ and $U_F'$, respectively (for simplicity, we assumed there is a single $U_F'$ that corresponds to $a=1$. Other cases like two different Floquet operators in [Fig.~\ref{fig:tensornetwork}(e)] can be considered in the same way). 
When $U_F'$ has a quasi-energy $\theta_j'$ satisfying $U_F'|\phi_j\rangle=e^{i\theta_j'}|\phi_j\rangle$,
\begin{align}
    \text{tr}\,{U_F'}^{t-t_0}U_F^{t_0}
    =
    \sum_{j}e^{i(t-t_0)\theta_j'}\langle\phi_j| U_F^{t_0}|\phi_j\rangle.
\end{align}
Compared with Eq.~(\ref{eq:sff}), each contribution of quasi-energy $\theta_j$ to the trace is reduced by a factor $\langle\phi_j| U_F^{t_0}|\phi_j\rangle$.
The absolute value of this factor is less than unity unless $|\phi_j\rangle$ is a simultaneous eigenstate of $U_F$ and $U_F'$.
This indicates that the reduction of the SFF at $t_0$ quantifies the mismatch of the eigenstates before and after $t_0$ or, in other words, 
the change of the wavefunctions by the twist.
The twist affects relevantly around the self-dual point [Fig.~\ref{fig:sff_boundarychange}(e)], while it is almost irrelevant at the borders of the phase diagram, which is deep in the many-body localized phase [Fig.~\ref{fig:sff_boundarychange}(h)].
Hence, our twist may not be used to characterize the other three (0SG, PM, and 0$\pi$PM) phases.

\section{Conclusion}
\label{sec:conclusion}

In this study, we investigated the response of the DTC phase in periodically driven Floquet systems to the time-translation twist.
Since DTCs spontaneously break the discrete time translation symmetry of the drive, they are sensitive to the boundary condition twisted by time translation, even though they are also many-body localized.
The sensitivity serves as a diagnostic for the spontaneously broken phase 
and is analogous to the Little-Parks effect for superconductors.

As a specific model, we considered the DTC phase of the one-dimensional disordered kicked Ising model.
In addition, we introduced two extensions of the time-translation twist.
One is to twist the boundary by a fraction of the driving period. 
The other applies the twist after some preparation time under periodic boundary conditions.
They disturb the formation of the DTC by 
creating a mismatch between
the phase of the Floquet drive and the system's oscillation across the boundaries, respectively.
The effect of the fractional-period twist is evident in the short-time behavior: the SFF with a half-period twist vanishes when the system's dominant oscillation period is twice the driving one.
On the other hand, the effect of applying the twist after a preparation time
appears in the long-time behavior: the plateau value of the SFF with a single-period twist is reduced largely compared with that with a double-period twist.
This reduction indicates that the single-period twist significantly changes the spin wavefunction due to the mismatched oscillation phase.
Our diagnosis of the long-time behavior is applicable to generic systems when the Floquet operator is represented as a matrix-product operator, while that of the short-time behavior requires space-time duality.

Compared with previous works on the DTC order \cite{PhysRevLett.116.250401,PhysRevLett.117.090402,PhysRevLett.118.030401},
one of the advantages of our method lies in the ability to evaluate the rigidity of the DTC order quantitatively.
This is analogous to the relation between the superconducting order and the U(1) twisted boundary condition, which inhibits the superconductors from having a coherent bulk U(1) phase. 
The rigidity is quantified by the variation of the partition function against the twist as shown in Eq.~(\ref{eq:meissner}).
A similar argument can be applied to the DTC with time-translation twist,
and hence the quantity we studied in Sec.~\ref{eq:long-time} can be regarded as the rigidity (stiffness) of the DTC order albeit the twist is discretized.
In addition, the effect of the twist persists even after the Heisenberg time, 
when the oscillation of the spins can no longer be observed.
Notice that our method does not depend on any specific mechanism of the ergodicity breaking necessary for the stability of DTCs.
Hence, we anticipate that our method is applicable to other DTCs like prethermal ones.

 As for future directions, one intriguing possibility is to explore the implementation of the time-translation twist
  in quantum devices. 
For example, the time-translation twist could be applicable to the DTC in quantum computers \cite{doi:10.1126/sciadv.abm7652} based on the tensor network representation in Fig.~\ref{fig:tensornetwork}.
On a more fundamental level, it would also be valuable to further investigate the theoretical role of the time-translation twist.
Notably, such a boundary condition can be interpreted as a time-like topological defect in spacetime. Recently, other forms of topological spacetime defects in Floquet systems have been studied in Refs.\ \onlinecite{10.21468/SciPostPhys.16.3.075, Kishony2025}, where the algebra of defects and Majorana-bound states has been discussed.

\begin{acknowledgments}
R.N. is supported in part by JSPS KAKENHI Grant No.~JP24K06926.
S.R.\ is supported
by a Simons Investigator Grant from the Simons Foundation (Award No.~566116).
\end{acknowledgments}

\appendix

\section{Vanishing SFF at $a=1/2$}

Here, we will show that the ratio $\langle K(t,a=1/2)\rangle/\langle K(t,a=0)\rangle$ between the half-twisted and untwisted averaged SFFs vanishes
when the leading Lyapunov exponent of the column-to-column transfer matrix is in the $\pi$ frequency subspace.

\subsection{$t=2$}
\label{app:t2}

First, we consider $t=2$.
We denote a state along the temporal direction by a $\tau^z$'s eigenstate $|\tau_1,\tau_2\rangle$ where $\tau_k^z|\tau_1,\tau_2\rangle=\tau_k|\tau_1,\tau_2\rangle$.
To implement a fractional twist by $a=p/q$, we decompose the Floquet operator into $q$ parts by Eq.~(\ref{eq:floquetunitary_decomposition}). The resulting state is written on $2q$ temporal spins.
Since $H_I$ is diagonal on the $\tau^z$ basis, the state is simplified as $|\tau_1,\cdots,\tau_1,\tau_2,\cdots,\tau_2\rangle$.
For brevity, we denote such a state as $|\tau_1,\tau_2\rangle$.
On the basis $(|\!\uparrow\uparrow\rangle,|\!\uparrow\downarrow\rangle,|\!\downarrow\uparrow\rangle,|\!\downarrow\downarrow\rangle)$, where $\uparrow/\downarrow$ should be identified with $+1/-1$, the column-to-column transfer matrix (\ref{eq:ctctransfermatrix}) from a site $j$ to $j+1$ is given by
\begin{align}
    \tilde{U}_{KI}^{(j)}
    =
    \begin{pmatrix}
        e^{-2iJ_j} & 1 & 1 & e^{2iJ_j}\\
        1 & e^{-2iJ_j} & e^{2iJ_j} & 1\\
        1 & e^{2iJ_j} & e^{-2iJ_j} & 1\\
        e^{2iJ_j} & 1 & 1 & e^{-2iJ_j}
    \end{pmatrix}
    T^z_j,
\end{align}
where $T^z_j=\text{diag}[e^{-2ih_j}\cos^2 b, -\sin^2 b, -\sin^2 b, e^{2ih_j}\cos^2 b]$
and $j\in[1,L-1]$.
When the boundary is twisted by $a=p/q$, the column-to-column transfer matrix across the boundary is
\begin{align}
    \tilde{U}_{KI}^{(L)}
    =
    \begin{pmatrix}
        e^{-2iJ_L} & 1 & 1 & e^{2iJ_L}\\
        1 & e^{-2i(q-2p)J_L/q} & e^{2i(q-2p)J_L/q} & 1\\
        1 & e^{2i(q-2p)J_L/q} & e^{-2i(q-2p)J_L/q} & 1\\
        e^{2iJ_L} & 1 & 1 & e^{-2iJ_L}
    \end{pmatrix}
    T^z_L,
\end{align}
while the bulk transfer matrix is unchanged.

As we mentioned in Sec.~\ref{sec:shorttime}, a state $|\pi\rangle=|\!\uparrow\downarrow\rangle-|\!\downarrow\uparrow\rangle$, which correspond to a vector $(0,1,-1,0)$, is an eigenstate of both the bulk and boundary column-to-column transfer matrix.
The corresponding eigenvalue is 
\begin{align}
    &\lambda_j=2i\sin2J_j\sin^2 b, \\
    &\lambda_L(q/p)=2i\sin[2(1-2p/q)J_L]\sin^2 b, 
\end{align}
for the bulk and boundary column-to-column transfer matrices, respectively.
It is clear that $\lambda_L$ vanishes when $a=q/p=1/2$.

When the leading Lyapunov exponent is in the $\omega=\pi$ subspace, the SFF under the periodic boundary condition is given by 
\begin{align}
    \langle K(2,0)\rangle=\left|\prod_j^{L-1}\lambda_j\cdot \lambda_L(0)\right|^2,
\end{align}
in a sufficiently large system.
On the other hand, when we twist by $a=1/2$, the contribution from the $\omega=\pi$ subspace vanishes since $\lambda_L(1/2)=0$ and the subleading contribution from the $\omega=0$ subspace becomes leading.
So, $\langle K(2,1/2)\rangle$ is approximately the subleading contribution to $\langle K(2,0)\rangle$.
Assuming there is a finite gap between leading and subleading Lyapunov exponents,
the ratio $\langle K(2,1/2)\rangle/\langle K(2,0)\rangle$ converges exponentially to zero as the system size increases.

In the clean limit $J_j=J$ and $h_j=h=0.6$, we checked that the region where the leading Lyapunov exponent is in the $\omega=\pi$ subspace [Fig.~\ref{fig:kickedising_sffhalf}(e)] agrees with that of vanishing ratio at $a=1/2$ [the dark region in Fig.~\ref{fig:kickedising_sffhalf}(a)].

\subsection{$t=4$}
\label{app:t4}

For $t=4$, the Hilbert space along the temporal direction on which the column-to-column transfer matrices act has $2^4$ dimensions.
The Hilbert space is decomposed by the frequency $\omega=2\pi n/4\,(n\in[0,3])$.
Let a state be denoted by $|\tau_1,\tau_2,\tau_3,\tau_4\rangle$ where $\tau_k^z|\tau_1,\tau_2,\tau_3,\tau_4\rangle=\tau_k|\tau_1,\tau_2,\tau_3,\tau_4\rangle$.
By the translation operator $\hat{T}$ defined by $\hat{T}|\tau_1,\tau_2,\tau_3,\tau_4\rangle=|\tau_4,\tau_1,\tau_2,\tau_3\rangle$,
eigenstates with frequency $\omega$ are given by 
\begin{align}
    |\omega\rangle\propto\sum_{p=0}^3e^{-i\omega p}\hat{T}^p|\tau_1,\tau_2,\tau_3,\tau_4\rangle
\end{align}
unless it vanishes.
Specifically, the $\omega=\pi$ subspace is spanned by 4 states, which are defined by
\begin{align}
    &|\pi_1\rangle\propto\sum_{p=0}^3(-1)^p\hat{T}^p|\uparrow,\downarrow,\uparrow,\downarrow\rangle,\\
    &|\pi_2\rangle\propto\sum_{p=0}^3(-1)^p\hat{T}^p|\uparrow,\uparrow,\uparrow,\downarrow\rangle,\\
    &|\pi_3\rangle\propto\sum_{p=0}^3(-1)^p\hat{T}^p|\uparrow,\downarrow,\downarrow,\downarrow\rangle,\\
    &|\pi_4\rangle\propto\sum_{p=0}^3(-1)^p\hat{T}^p|\uparrow,\uparrow,\downarrow,\downarrow\rangle.
\end{align}
Here, a unitary transformation is applied to make a part of the basis states renewed as
\begin{align}
    \begin{pmatrix}
        |\pi_2'\rangle \\
        |\pi_3'\rangle
    \end{pmatrix}
    =
    \frac{1}{\sqrt{2}}
    \begin{pmatrix}
        e^{2ih_L} & e^{-2ih_L} \\
        e^{2ih_L} & -e^{-2ih_L}
    \end{pmatrix}
    \begin{pmatrix}
        |\pi_2\rangle \\
        |\pi_3\rangle
    \end{pmatrix}.
\end{align}
In the following, we use these new states (we omit primes) together with the original $|\pi_1\rangle$ and $|\pi_4\rangle$.

On this basis, the column-to-column transfer matrix across the boundary with the twist by $a=1/2$ is given by
\begin{align}
    &\langle \pi_k|\tilde{U}_{KI}^{(L)}|\pi_l\rangle
    =
    \sqrt{2}\sin^2 2b \sin^2 J_L
    \begin{pmatrix}
        0 & 0 & 0 & 0\\
        0 & 0 & 0 & i\sin 2h_L \\
        0 & 0 & 0 & -\cos 2h_L \\
        0 & 0 & 1 & 0
    \end{pmatrix}_{kl}.
\end{align}
Therefore, $|\pi_1\rangle$ and $|\pi_2\rangle$ are null states of the boundary column-to-column transfer matrix.
Since the diagonal element is zero even for $|\pi_3\rangle$ and $|\pi_4\rangle$,
a linear combination of $|\pi_1\rangle$, $|\pi_2\rangle$, and $|\pi_3\rangle$, or that of $|\pi_1\rangle$, $|\pi_2\rangle$ and $|\pi_4\rangle$ has a vanishing expectation value of the boundary column-to-column transfer matrix.

The column-to-column bulk transfer matrix is block-diagonalized as
\begin{align}
    &\langle \pi_k|\tilde{U}_{KI}^{(j)}|\pi_l\rangle
    =
    -4\sin^2 b\sin 2J_j \times\notag\\
    &
    \begin{pmatrix}
        i\cos 2J_j & -i & 0 & 0\\
        i\cos 2h_j & -i\cos 2J_j\cos 2h_j & i\sin 2h_j & 0 \\
        \sin 2h_j & -\cos 2J_j\sin 2h_j & -\cos 2h_j & 0 \\
        0 & 0 & 0 & -1 
    \end{pmatrix}_{kl}
    A_j,
\end{align}
where $A_j=\text{diag}[\sin^2 b,\cos^2 b,\cos^2 b\sin 2J_j,\sin^2 b\sin 2J_j]$.
Therefore the eigenstate of the bulk column-to-column bulk transfer matrix is $|\pi_4\rangle$ and 
linear combinations of $|\pi_1\rangle$, $|\pi_2\rangle$, and $|\pi_3\rangle$.
Since all the states have vanishing expectation value of the boundary column-to-column 
transfer matrix, 
one can conclude that the ratio at $t=4$ vanishes in a sufficiently large system when the leading Lyapunov exponent is in the $\omega=\pi$ subspace.

Unlike the $t=2$ case, not all the states in the $\omega=\pi$ subspace are null states of the boundary column-to-column bulk transfer.
Thus, we need to rely on the bulk property.
For this reason, it may not be easy to extend the relation between vanishing ratio at $a=1/2$ and the DTC to arbitrary $t$.

Notice that a region around $\bar{J}=0$ and $b\simeq 3\pi/8$ in Fig.~\ref{fig:kickedising_sffhalf}(b) also has a vanishing SFF at a half twist.
This region is considered to be resulting from two DTC regions in $\bar{J}>0$ and $\bar{J}<0$ in Fig.~\ref{fig:kickedising_sffhalf}(f) due to symmetry with respect to $\bar{J}\to -\bar{J}$.

\section{Long-time behavior with $t_0=0$}
\label{app:t00}

\begin{figure}
    \centering
    \includegraphics[width=0.49\textwidth]{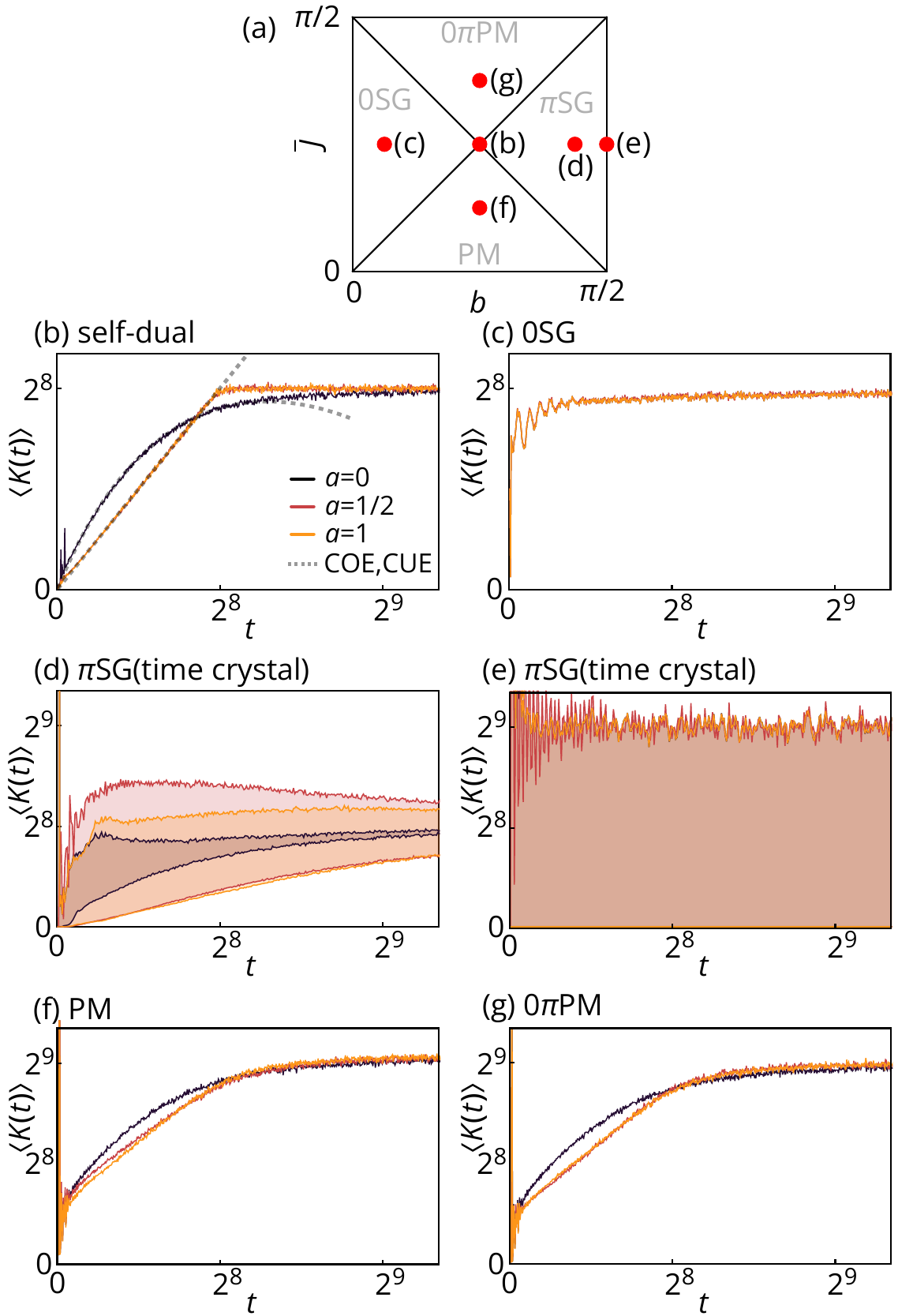}
    \caption{
    (a) The phase diagram of the kicked Ising model. 
    (b)-(g) The SFF averaged over 10000 samples with length $L=8$, longitudinal field mean $\bar{h}=0.6$, variance $\pi/10$, and Ising coupling variance $\pi/10$.
    Each figure corresponds to
    (b) the self-dual point $(b,\bar{J})=(\pi/4,\pi/4)$,
        (c) the 0SG phase $(b,\bar{J})=(\pi/16,\pi/4)$,
    (d)-(e) the $\pi$SG (DTC) phase ($b,\bar{J})=(7\pi/16,\pi/4)$ and $(b,\bar{J})=(\pi/2,\pi/4)$,
    (f) the PM phase $(b,\bar{J})=(\pi/4,\pi/8)$,
    and (g) the 0$\pi$PM phase $(b,\bar{J})=(\pi/4,3\pi/8)$, respectively.
    In each figure, the SFF without twist (black), with a half twist $a=1/2$ (purple) and with a single twist $a=1$ (orange) are shown. 
    In (b), the SFF of the random matrix theory (COE and CUE) is shown by dashed lines.
    In (d) and (e), the SFF at even $t$ and odd $t$ are shown independently by solid lines while lightly colored regions represent the oscillation.
        \label{fig:kickedising_sff-t}
    }
\end{figure}

Here we discuss the averaged SFF when the time-translation twist
is imposed from the outset. 
We numerically evaluate the time-dependence of the SFF of the kicked Ising model of length $L=8$ with and without the twist (Fig.~\ref{fig:kickedising_sff-t}).
We used twist parameter $a=0, 1/2$ and $1$, and the other parameters are the same as those in the main text.

The conclusion here is that the SFF is almost insensitive to the twist [Fig.~\ref{fig:kickedising_sff-t}(c),(d),(e),(f),(g)] except at the self-dual point [Fig.~\ref{fig:kickedising_sff-t}(b)],
and that the formation of DTC is not disturbed by the twist [Fig.~\ref{fig:kickedising_sff-t}(d),(e)].
The insensitivity is considered to be due to the many-body localization.
The DTC oscillation inside the $\pi$SG phase is rather enhanced by the twist, where another form of the DTC, different from the one under PBC, is formed under the twist.

Some comments are in order.
(i) At the self-dual point [Fig.~\ref{fig:kickedising_sff-t}(b)], the growth up to the Heisenberg time $t=2^L$ changes from that of COE to that of CUE [see Eq.~(\ref{eq:spectralformfactor_randommatrixtheory})] by the twist as time-reversal symmetry is broken.
(ii) The long-time behavior is symmetric with respect to $\bar{J}=\pi/4$.
(iii) On the line $b=\pi/2$ (in the DTC phase), the SFF shows the DTC oscillation between 0 and $2^{L+1}$  [Fig.~\ref{fig:kickedising_sff-t}(e)], which is explained analytically in Appendix \ref{sec:bpi2}.

\section{$b=\pi/2$}
\label{sec:bpi2}

We confirm two features of the SFF at $b=\pi/2$, that is, the long-time behavior of the SFF is the oscillation between 0 and $2^{L+1}$,
and the ratio between untwisted and half-period twisted SFFs vanishes in a short time.

Along the line $b=\pi/2$, the kick part of the Floquet operator reduces to the spin-flip operator $P=\prod_j\sigma_j^x$.
As a result, the SFF vanishes at odd $t$, and the longitudinal field is canceled after even periods as it anticommutes with $P$.
In the following, we focus on even $t$.
Let the eigenvalue of $\sigma_j^z$ be denoted by $s_j$.
Then,
\begin{align}
    \text{tr}[U_{KI}^t]
    =
    \sum_{\{s_j\}}
    \exp\left[-it\sum_{j=1}^LJ_js_js_{j+1}\right].
    \label{eq:tracefloquet_untwisted_main}
\end{align}
After averaging over the Ising coupling and tracing over spins, the SFF without twist becomes
\begin{align}
    &\langle K(t)\rangle
    =
    2^L\left[
        \left((-i)^L+i^L\right)e^{-2Lt^2\sigma_J^2}\sin^L 2t\bar{J}\right.\notag\\
        &\left.+\left(1+e^{-2t^2\sigma_J^2}\cos 2t\bar{J}\right)^L
        +\left(1-e^{-2t^2\sigma_J^2}\cos 2t\bar{J}\right)^L
    \right],
\end{align}
where, only in this section, we use the average over the Gaussian distribution
\begin{align}
    \langle A[\bm{J}]\rangle
    =
    \int_{-\infty}^\infty \prod_j\frac{d J}{\sqrt{2\pi}\sigma_J}
    e^{-(J_j-\bar{J})/2\sigma_J^2}A[\bm{J}]
\end{align}
with mean $\bar{J}$ and variance $\sigma_J$.
Typically, the averaged SFF oscillates between 0 and $\sim 2^{2L}$ for a short time ($t\sigma_J\ll 1$) while between 0 and $\sim 2^{L+1}$ for a long time ($t\sigma_J\gg 1$) .

With the time-translation twist,
the right-most spin is coupled to the left-most one at a different time.
Let us begin with the twist by an integer $a$.
The Ising term across the boundary couples $s_L$ at time $\tau$ and $s_1$ at time $\tau+a$, which is $(-1)^a s_1$ as all spins alternate by $P^a$. 
As a result, the time-translation twist
modifies a part of the exponent in Eq.~(\ref{eq:tracefloquet_untwisted_main}) as
\begin{align}
    J_Ls_Ls_1
    \to
    (-1)^aJ_Ls_Ls_1.
    \label{eq:isingcoupling_integertwist_main}
\end{align}

This procedure can be generalized to a fractional twist $a=p/q$ where $p\in[0,q)$ for a moment, for simplicity.
After the decomposition of the Floquet operator by Eq.~(\ref{eq:floquetunitary_decomposition}), 
there are $q$ rows in a single period in the tensor network [Fig.~\ref{fig:tensornetwork}(c)].
Most of the Floquet operators contain only $\sigma^z$ while the last one $U_{(q-1)F}$ contains $P$.
Thus, the boundary states $s_1$ and $s_L$ are unchanged from $t=0$ to $t=(q-1)/q$.
After shifting the boundary [Fig.~\ref{fig:tensornetwork}(c)], there are $p$ rows out of $q$ that couple $s_L$ and $-s_1$, while the other $q-p$ rows couple $s_L$ and $s_1$.
The resulting Ising term interpolates Eq.~(\ref{eq:isingcoupling_integertwist_main}) as
\begin{align}
    J_Ls_Ls_1
    \to
    \frac{q-2p}{q}J_Ls_Ls_1
    =
    (1-2a)J_Ls_Ls_1.
    \label{eq:isingcoupling_fractionaltwist}
\end{align}
This indicates that $a=1/2$ is a special case where the Ising coupling across the boundary is severed by the twist.

The averaged SFF after a fractional twist $a=p/q$ is then given by
\begin{widetext}
\begin{align}
    \langle K(t;a)\rangle
    =
    2^L&\left[
        \left((-i)^L+i^L\right)\exp\left[-2t^2\sigma_J^2\left(L-1+(1-2a)^2\right)\right]\sin^{L-1} 2t\bar{J}\sin \left[2t\bar{J}(1-2a)\right]\right.\notag\\
        &\left.+\sum_{\pm}\left(1\pm e^{-2t^2\sigma_J^2}\cos 2t\bar{J}\right)^{L-1}\left(1\pm\exp\left[-2t^2\sigma_J^2(1-2a)^2\right]\cos \left[2t\bar{J}(1-2a)\right]\right)
    \right].
    \label{eq:sff_bpi2}
\end{align}
\end{widetext}
The ratio in a sufficiently large system (typically $L\gg e^{2t^2\sigma_J^2}$) reduces to
\begin{align}
    \frac{\langle K(t;a)\rangle}{\langle K(t;0)\rangle}
    \to
    \frac{1\pm\exp\left[-2t^2\sigma_J^2(1-2a)^2\right]\cos \left[2t\bar{J}(1-2a)\right]}{1\pm\exp\left[-2t^2\sigma_J^2\right]\cos 2t\bar{J}},
    \label{eq:ratiovanish_app}
\end{align}
where $\pm$ corresponds to the case $\cos 2t\bar{J}\gtrless 0$, respectively.
The ratio vanishes at $a=1/2$ when $\cos 2t\bar{J}< 0$. 
Specifically for $t=2$, $\cos 2t\bar{J}< 0$ indicates $\pi/8<\bar{J}<3\pi/8$, which agrees with Fig.~\ref{fig:kickedising_sffhalf}(a) at $b=\pi/2$.

Some comments are in order.
(i) After a long time $t\sigma_J\gg 1$, the SFF at $b=\pi/2$ becomes insensitive to the time-translation twist
since Eq.~(\ref{eq:sff_bpi2}) reduces to $2^{L+1}$.
(ii) $a=1/2$ is a point where the mismatch of the drive phase is maximized.
That is, the boundary spins are kicked alternately, distorting the DTC formation.

\bibliography{twistFloquet.bbl}

\end{document}